\documentclass{PoS}
\usepackage[authoryear,square]{natbib}
\bibpunct{(}{)}{;}{a}{}{,}

\title{Maser Astrometry with VLBI and the SKA}

\ShortTitle{Maser astrometry}

\author{
\speaker{James A. Green}$^1$,
Huib Jan van Langevelde$^2$, 
Andreas Brunthaler$^3$, 
Simon Ellingsen$^4$, 
Hiroshi Imai$^5$, 
Wouter Vlemmings$^6$,  
Mark Reid$^7$, 
and Anita Richards$^8$  
\\
$^1$SKA Organisation
$^2$JIVE/Sterrewacht Leiden
$^3$MPIfR Bonn
$^4$University of Tasmania
$^5$Kagoshima University
$^6$Chalmers University of Technology
$^7$Harvard-Smithsonian CfA
$^8$University of Manchester
\\
E-mail: \email{J.Green@skatelescope.org}
}

\abstract{
We discuss the unique opportunities for maser astrometry with the inclusion of the Square Kilometre Array (SKA) in Very Long Baseline Interferometry (VLBI) networks. The first phase of the SKA will enable observations of hydroxyl and methanol masers, positioning the latter to an accuracy of 5 microarcseconds, and the second phase may allow water maser observations. These observations will provide trigonometric distances with errors as small as 1\%. The unrivalled sensitivity of the SKA will enable large-scale surveys and, through joint operations, will turn any VLBI network into a fast astrometry device. Both evolved stars and high mass star formation regions will be accessible throughout the (Southern) Milky Way, completing our understanding of the content, dynamics and history of our Galaxy. Maser velocities and proper motions will be measurable in the Local Group of  galaxies and beyond, providing new insights into their kinematics and evolution.
}

\FullConference{
Advancing Astrophysics with the Square Kilometre Array\\
June 8-13, 2014\\
Giardini Naxos, Italy}

\begin{document}

\section{Introduction}
Through its sheer collecting area, the Square Kilometre Array (SKA) will have unprecedented sensitivity for spectral lines. The full receiver suite of the first phase of the SKA (SKA1) will include the hydroxyl (OH) ground- and excited-state maser transitions (at 1.612, 1.665, 1.667, 1.720, 4.765, 6.030 and 6.035\,GHz), methanol (CH$_{3}$OH) maser transitions (at 6.668 and 12.178\,GHz), and other rarer maser transitions such as from the formaldehyde (H$_{2}$CO) molecule (at 4.8\,GHz). The second phase of the SKA (SKA2) has the prospect to allow for studies of water (H$_{2}$O) masers (at 22.235\,GHz). This frequency coverage, in joint operation with one or more Very Long Baseline Interferometry (VLBI) network(s) \citep{paragi}, will allow revolutionary science to be done in the field of Galactic and extragalactic structure and dynamics during all phases of the SKA.

In this chapter we highlight the VLBI capabilities for maser astrometry of SKA1-MID and SKA1-SUR, combined with existing telescopes in the Southern Hemisphere (the Long Baseline Array, LBA, in Australia, New Zealand and South Africa), Asia (including VLBI networks in Japan, South Korea and China), North America (the Very Long Baseline Array, VLBA) and Europe (the European VLBI Network, EVN). We will assume this will be complemented by further development of the various VLBI arrays, notably in Africa (the African VLBI Network, AVN), and potentially in South America as well.  This will be an array with competitive uv-coverage and incredible sensitivity, dominated by the collecting area of the SKA element(s). As the SKA will synthesise a small beam on the sky, for the most efficient observing (and to reach the highest accuracies), it will be necessary to form multiple beams with the SKA to cover both targets and calibrators (position reference sources) within a single beam of the other VLBI telescopes \citep{paragi}.

Maser astrometry yields a direct measurement of the distance and proper motions of individual young and evolved stars,  calibrating their astrophysical properties. Moreover, major strides have been made in recent years in our understanding of the Milky Way through maser astrometry (see \citealt{reid14a} and \citealt{reid14b} for recent compilations and analysis). The developments include: refining parameters of Galactic rotation and the local standard of rest; refining the distance to the Galactic centre; and establishing accurate configurations of the spiral arms. These developments have wide ranging impacts, such as in the testing of general relativity through the effect of gravitational radiation damping binary pulsar systems (where the accuracy of the solar distance and rotation are fundamental, \citealt{reid14a}).

The timeline for the first phase of the SKA will overlap with the expected timescale for the results of the {\it Gaia} mission. The radio astrometry of VLBI with the SKA and the optical astrometry of {\it Gaia} will be highly complementary, with the former providing distances to the dust enshrouded regions and individual evolved stars, the latter the formed and optically radiant stars. This is particularly true for the inner Galaxy, where {\it Gaia}'s ability to penetrate the dust is limited. 

\begin{figure}
\begin{center}
\renewcommand{\baselinestretch}{1.1}
\includegraphics[width=9cm]{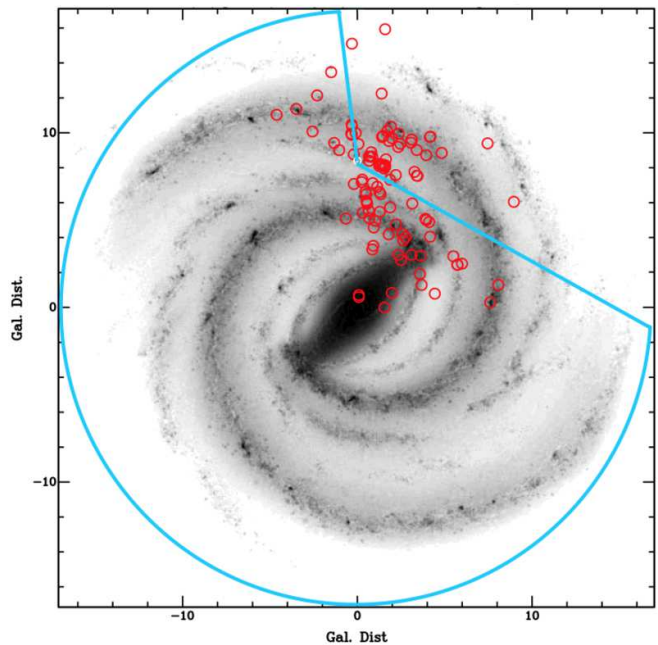}  
\caption{\small Masers (red circles) with astrometric parallaxes from the BeSSel survey and other observations \citep{reid14a}, overlaid on the informed artist impression of the Milky Way (R. Hurt: NASA/JPL- Caltech/SSC). The blue lines enclose the regions accessible with Southern Hemisphere SKA-VLBI, in which there are of the order of 1000 known methanol masers.}
\label{FluxRatios}
\end{center}
\end{figure}

\section{The technique of VLBI maser astrometry}
As an introduction to the science enabled by VLBI we briefly outline the technique of maser astrometry: it utilises position monitoring measurements of the maser source relative to an appropriate distant calibrator (background quasar), precisely determining a proper motion and a position (through parallax) which gives the distance to the source. Phase-referencing astrometric observations involving a background quasar and a maser source essentially involve measuring the phase delay between the two sources to determine their relative positions.  This requires precise calibration of all the factors which contribute to the observed delay, including the station and source coordinates, instrumental and atmospheric delays (see \citealt{honma08} for an in depth discussion, and Section 5 of this paper for discussion in the context of SKA). The following sections describe the Galactic and extragalactic science achievable through this technique.

\section{Aims of astrometric measurements for Galactic science}
The crucial scientific goal of maser astrometry with the SKA is obtaining distances to hundreds of sources in the Southern sky, together with a significant increase in Northern sources accessible with the SKA (south of declination 20$^{\circ}$), more than doubling those obtained to date. We will simultaneously obtain the full three dimensional kinematics, through the combination of the line-of-sight velocity (from the Doppler shift of the line) and the proper motion. With the deployment of band 5 receivers on SKA1-MID (as defined in \citealt{dewdney13}), the widespread 6.7 and 12.2\,GHz methanol masers, proven to be exclusive probes of high mass star formation \citep{breen13}, will be studied.  Due to their tight association with the spiral arms and molecular gas, they directly trace the spatial structure and dynamics of the Galaxy. To date over 100 maser parallaxes have been measured in the Northern Hemisphere, through the Be{SS}el project \citep{brunthaler11}, together with observations with the Japanese VLBI Exploration of Radio Astronomy, VERA, network. With Southern Hemisphere VLBI including the SKA, the current results can be improved through several hundred more maser measurements, providing a total sample of $\sim$1000, reaching an order of magnitude better statistics over the current sample. These observations will complement the {\it Gaia} results, for which close to the Galactic plane, can only penetrate limited distances. Together with radio continuum astrometry, we have the capability to measure the distance to the majority of star forming regions that the Atacama Large Millimeter Array (ALMA) can observe. The precise distances (1\% errors for distances of $\sim$2\,kpc, rising to 5\% for $\sim$10\,kpc) will allow the derivation of accurate physical parameters (including linear scale, mass, luminosity) and will offer the opportunity to improve distances to associated objects (including star forming regions, supernova remnants, planetary nebulae). Besides a handle on the dark matter distribution, the formation and evolution of the Galaxy can be studied through its dynamics and the orbital motions of the Magellanic Clouds together with signatures of past and recent mergers (see Section 4).

VLBI capabilities for the hydroxyl and methanol maser lines also enable detailed studies of the gas kinematics and excitation on the scales at which accretion processes in high-mass stars are expected to occur (through the `3-D' velocities provided by the line-of-sight velocities and proper motions). This will complement ALMA observations which have more modest resolution, but are sensitive to thermal excitations.

Together with atomic hydrogen, SKA1-MID and SKA1-SUR will be sensitive to the hydroxyl main and satellite lines at 18cm (each with their respective band 2 receivers, as defined in \citealt{dewdney13}). These lines can be used to trace star forming regions, the circumstellar envelopes of evolved stars and supernova remnants. Compared to previous efforts, the SKA will have the sensitivity to quickly survey the Galaxy \citep{etoka}. However, the moderate resolution of the SKA alone will leave some confusion when comparing with objects at other wavelengths. As such, VLBI baselines that include SKA1-MID and SKA1-SUR will be required. The existing LBA and EVN are both well equipped in this band and the AVN will benefit from this frequency capability.

Astrometry of circumstellar OH masers can yield proper motions and parallaxes of individual AGB stars \citep{vlemmings07b}. Compared to methanol masers the method is limited by the poorer resolution and the intrinsically limited brightness of the OH masers. However with the SKA in a VLBI array very many objects within a few kpc should be accessible. The direct measurements of distance are fundamental for quantitatively understanding the physics and chemistry processes associated with these mass-losing stars, calibrating their feedback to the Galactic lifecycle of elements. Measuring distances to highly obscured stars and in highly obscured parts of the Galaxy is again complementary to what {\it Gaia} will deliver. Radio observations can for example extend determinations of the Period-Luminosity relation for AGB stars to the regime with higher mass-loss by including more extreme objects. When the capabilities of SKA and the VLBI array cover water masers, the higher frequencies will enable us to overcome the effects of scintillation and it will become feasible to sample stellar orbits in the inner Galaxy, in which the dynamic signature of the bar and merging events could be detectable. 

In addition to regions of star formation and evolved objects, the satellite line transition at 1720 MHz is found in supernova remnants, and as such is used to probe the interaction with surrounding molecular clouds and examine the processes involved \citep[e.g.][]{brogan13}. Maser astrometry of this transition will demonstrate the three dimensional geometry, provide proper motions for analysis of turbulent and shocked structures, and precisely locate the masers relative to x- and gamma-ray radiation (enabling tests of pre- and post-shocked conditions).

Furthermore, astrometric measurements will complement the study of the large-scale Galactic magnetic field through Zeeman measurements of hydroxyl masers, tying Galactic structure determined through parallax with the fields, and the study of small-scale fields around young massive stellar objects through Zeeman measurements of methanol masers \citep{robishaw}.

\subsection{Phased deployment of SKA and expected results}
Maser astrometry with VLBI (using `traditional' phase referencing) represents a field which has the potential for scientific return through utilising antennas in early phase SKA (i.e. in the build up to phase 1, when there are less than 50\% of the antennas available). An example would be incorporating SKA1-SUR antennas with the LBA, where even a few dishes will make a measurable difference with the benefit of enhanced east--west baselines. The first science from the Australian SKA Pathfinder (ASKAP) was a VLBI experiment with one ASKAP antenna included in the LBA \citep{tzioumis10}.

For the first phase of the SKA, we can expect System Equivalent Flux Densities (SEFDs) of 2.1\,Jy at the hydroxyl frequencies and 2.8\,Jy at the methanol frequency \citep{dewdney13}. There are currently $\sim$1200 known sources of methanol maser emission \citep{pestalozzi05,green09a}. Of the order of 300--400 of these will be observed from the Northern Hemisphere with existing VLBI networks \citep[e.g.][]{brunthaler11}, and we can expect, through $\sim$1500 hrs observing, to detect and obtain 5 microarcsecond parallaxes for $\sim$300 from the Southern Hemisphere with SKA1. This would allow observations of methanol masers with peak flux densities $\ge$0.7\,Jy, more than an order of magnitude weaker than can currently be observed with the LBA. SKA1 will also be able to detect several thousand hydroxyl maser sources \citep{japaneseSKA} with SKA1-MID and SKA1-SUR combined, although not all of these will be accessible for direct parallax measurements because of the limited resolution at frequencies of $\sim$1.6\,GHz.

SKA2, with an expected sensitivity 10 times that of SKA1-MID and with extended baselines, will vastly improve on SKA1, with an expectation of being able to detect the very faintest methanol masers, a further 400-600 sources, bringing the total with parallax measurements to $\sim$1200. It has been suggested that the luminosity of methanol masers increases with age \citep{ellingsen13}, and the full SKA will allow direct testing of this hypothesis (as the less luminous masers should be found preferentially towards the leading edge of the spiral arms). Several thousand hydroxyl masers will be detectable with SKA2 \citep{etoka}.  Although not covered by the band designations in the current baseline design for SKA1, the dishes of SKA1-MID are able to work at 22 GHz, and SKA2 may be equipped with appropriate receivers. The SKA would then be sensitive to water masers from both evolved stars and young stellar objects. As water masers are intrinsically very bright, the capability to observe these sources has exciting applications, both in Galactic and extragalactic astrometry (see following Sections).

\section{Aims of maser astrometry for extragalactic science}
The SKA will enable us to significantly develop resolved stellar kinematic studies in other galaxies, in particular the Local Group of galaxies (the satellite galaxies of the Milky Way and beyond), enabling determination of their rotation, distances and proper motions. The kinematics of these galaxies are quite unique, providing an account of the history of the Milky Way including past and future interactions, insight into exotic events such as star bursts (as found in the Large Magellanic Cloud, e.g. \citealt{livanou06,anderson14}) and the dynamical friction in the galaxies. Hydroxyl, methanol and water masers can be used for this purpose, providing the opportunity for proper motion measurements \citep[for example as described in][]{dickey13}.  

The two nearest local group galaxies, the Large Magellanic Cloud (LMC) and the Small Magellanic Cloud (SMC), are prominent Southern Hemisphere objects.   At distances of $\sim50$\,kpc and $\sim62$\,kpc for the LMC and SMC respectively, they are a benchmark for studies of a variety of topics, including stellar populations, the interstellar medium, and the cosmological distance scale. Clearly evident in the radio are signs of interaction between the two individual Clouds \citep{putman03,bruns05}, and between the pair and the Milky Way \citep{mccluregriffiths08,nidever08,besla12}.  The Magellanic Clouds offer the best opportunity to view cosmological processes of hierarchical structure formation in action. Observations of the Magellanic Clouds can be compared with cosmological simulations to determine how interactions may have led to the triggering of star formation \citep{anderson14}. Stellar population studies have highlighted several episodes of heightened star formation at defined periods 0.2, 2, and 5\,Gyrs ago, raising the question as to how the timing and strength of these episodes may correspond with tidal interactions in the three body system.  

The SKA will have the sensitivity to make astrometric observations of significantly more maser sources in the LMC, SMC and other Local Group galaxies than can be achieved with current facilities.  These observations will enable more rapid and accurate determination of the proper motions of interstellar masers in these galaxies, and, combined with the line-of-sight component from the Doppler shift of the maser line, will provide full `3-D' velocities. The observed proper motions have three contributions: a component due to the proper motion of the Centre of Mass of the galaxy;
a component due to the orbital motion of the maser associated source about the Centre of Mass of the galaxy; 
a component due to internal motions of the masing gas within the star formation region. The first of these is the critical one for investigations of galaxy interaction, but the uncertainty in its determination depends upon both the precision in the proper motion measurement and the accuracy with which the other two components can be estimated. Observations of a larger number of individual maser regions in each southern local group galaxy will enable more accurate estimates of the proper motion of their centres of mass and also the orbital properties of the maser-associated objects.  These results will dramatically improve constraints on the physical processes of galaxy interaction by providing an accurate description of their motion.

Beyond the Magellanic Clouds, current VLBI arrays already have the astrometric accuracy to measure proper motions. However, the sensitivity of current arrays has limited this to only a handful of sources in nearby galaxies (for example M33, M31, and IC\,10). The superior sensitivity of the SKA (although in some cases at its northern limit) will allow us not only to detect many more masers within the Local Group, but also in nearby galaxy groups and even the Virgo cluster. This will allow us to understand the flow of galaxies in the local universe.

It is widely believed that the motion of the Milky Way relative to the cosmic microwave background (CMB), which is of the order of 500 km/s, is induced by mass concentrations within 150\,Mpc of the local universe, but there is a discrepancy between the direction of the motion and the distribution of visible mass in the local universe (see e.g. \citealt{loeb08} and references therein). This could be caused for example by a significant proper motion of the Milky Way relative to M31 or by a nearby structure in the Zone of Avoidance (behind the plane of the Milky Way). A measurement or a constraint on the proper motion of the Milky Way relative to the Virgo Cluster would be an important experiment to help understand this apparent discrepancy.

\subsection{Phased deployment of SKA and expected results}
Although the initial phases of SKA1 (with $\le$50\% of the antennas) will be beneficial for Galactic science, capabilities for extragalactic studies will only be enhanced over current capabilities once sensitivities approach those of the full deployment of SKA1 (i.e. SEFDs of 2.1 and 2.8\,Jy for the hydroxyl and methanol frequencies respectively). We expect that the GASKAP project \citep{dickey13} will detect around 100 hydroxyl maser in the Magellanic clouds, nearly an order of magnitude increase over the number currently known.  VLBI astrometry including SKA1 will enable their proper motions to be measured and significantly improve the accuracy to which the Centre-of-Mass motions are currently known (e.g. \cite{Kallivayalil13}). The strength of the full SKA will be in allowing kinematic measurements of galaxies beyond the Magellanic Clouds (as described in the previous Section).

\section{The requirements for VLBI maser astrometry with the SKA}
The general requirements for VLBI with the SKA are detailed in \citet{paragi}, but here we highlight specific requirements relevant for astrometric maser measurements, both in the `traditional' sense of phase referencing and in the `multi--view' sense of in-beam calibration.  The antenna positions must be accurately known (to a precision of a few millimetres) and there must be adequate calibrator sources with precise positions (positional accuracies of $\le$1\,milliarcsecond when used for removing the differential atmospheric delay residual). Once the coordinates (positions) of the masers are determined to an accuracy of a few tens of milliarcseconds, then the correlator will be required to average over a few seconds between outputs. This will cover position offsets of the individual maser spots from the determined source coordinates, taking into account time-average smearing. 

There will need to be the ability to determine zenith atmospheric delay contributions from the ionosphere and the troposphere. Traditionally rapid switching between the calibrator and target sources is used to effectively remove the short term atmospheric variations, however, it cannot estimate the delay difference due to the slightly different path through the airmass between the calibrator and target source. There are two components to the atmospheric delay, one due to the ionosphere which is dispersive and dominates at low frequencies and the other due to the troposphere which is non-dispersive and dominates at frequencies above 10\,GHz.  For SKA1 the ionosphere will be the dominant source of zenith delay uncertainty. Global Positioning Satellite (GPS) observations are routinely used to produce global ionospheric models, which can be used to reduce the impact of ionospheric effects by factors of two to five \citep{walker99}.  This technique yields residual zenith ionospheric delays accurate to around 3\,cm and it will be important to develop improved techniques, such as utilising continuous GPS observations at the telescope sites (e.g. \citealt{honma08}), to maximise the improved astrometry the SKA will facilitate. As described in \citet{paragi}, VLBI with the SKA will utilise multiple beams for calibration, which will aid the atmospheric calibration with in-beam phase referencing, providing an alternative to the `traditional' process described above, significantly reducing overheads (which often exceed 50\%). The experiments described in Sections 3 and 4 assume this `multi-view' method of observing. This requires adequate background calibrators (quasars) within the primary beam of the SKA antennas. At 1.6 GHz, based on the predictions of \citet{fomalont91} and \citet{condon07} we would expect plenty of sources, of the order of 450 sources brighter than 400\,$\mu$Jy, but at 6.7\,GHz this will be challenging in the first phase, with only $\sim$5 sources brighter than 100\,$\mu$Jy. The `multi-view' approach is also heavily dependent on the field of view of the other antennas within the VLBI array --  if the array includes antennas with diameters significantly exceeding the 15m of the SKA antennas, the number of available in-beam calibrators may be limited (the increased sensitivity is offset by the decreased field of view). However, a number of developments, such as multi-beam feeds or focal plane arrays for the larger antennas, and different calibration techniques, such as `boot strapping' and reverse phase referencing, may provide an avenue to overcome these limitations. 

For the hydroxyl and methanol observations, scheduling should be optimised to observe maxima in the parallax sinusoid, typically requiring four epochs over a one--to--two year period, for water maser observations more frequent scheduling will be required. However, the availability for VLBI in a sub array will allow appropriate scheduling, possibly concurrently with other observations (sub-arraying will additionally allow different size scales to be probed).  From a correlator perspective, VLBI maser astrometry will require wide bandwidths for continuum calibration and narrower bands with high spectral resolution for  the target maser emission. For the latter, a few tenths of km\,s$^{-1}$ or less  will be required for Galactic sources (necessary to sufficiently sample the maser spectrum and enable reliable identification of maser emission between epochs), and up to a few tens km\,s$^{-1}$ for the fainter extra-galactic OH masers.  Additionally, it is crucial to obtain a very high dynamic range (or signal-to-noise ratio) in the VLBI data, to achieve a sufficiently high astrometric accuracy for the ground-state hydroxyl transitions at $\sim$1.6--1.7\,GHz. In comparison for methanol, a dynamic range of 600-6000:1 is required for the maser source, 60:1 for the continuum calibrator. 

\section{Conclusions}

The SKA offers exciting options to study the dynamics of the Milky Way and nearby galaxies out to the Virgo Cluster. Maser astrometry offers a scientific return across the deployment of the SKA, from early stages through to the full SKA capability. Such studies will have important synergies with {\it Gaia} and ALMA in unraveling the spiral structure of the inner Galaxy, its dynamics and even its formation history. 

\setlength{\bibsep}{0.0pt}
\bibliographystyle{apj}
\bibliography{maservlbi.bib}

\end{document}